\documentclass[12pt]{iopart}

\usepackage{iopams} 
\usepackage{bm}
\usepackage{graphicx}
\begin{document}

\title[Finite-temperature Screening and the Specific Heat of Doped Graphene Sheets]{Finite-temperature Screening and the Specific Heat of Doped Graphene Sheets}

\author{M.R. Ramezanali}
\address{Department of Physics, Sharif University of Technology,
Tehran 11155-9161, Iran}
\author{M.M. Vazifeh}
\address{Department of Physics, Sharif University of Technology,
Tehran 11155-9161, Iran}
\author{Reza Asgari}
\ead{asgari@theory.ipm.ac.ir}
\address{School of Physics, Institute for research in fundamental sciences, IPM 19395-5531 Tehran, Iran}
\author{Marco Polini}
\ead{m.polini@sns.it}
\address{NEST-CNR-INFM and Scuola Normale Superiore, I-56126 Pisa, Italy}
\author{A.H. MacDonald}
\address{Department of Physics, The University of Texas at Austin, Austin, Texas 78712, USA}

\begin{abstract}
At low energies, electrons in doped graphene sheets are described by a massless Dirac fermion Hamiltonian. 
In this work we present a semi-analytical expression for the dynamical 
density-density linear-response function of noninteracting 
massless Dirac fermions (the so-called ``Lindhard"  function) at finite temperature.
This result is crucial to describe finite-temperature screening of interacting massless Dirac fermions within the Random Phase Approximation. 
In particular, we use it to make quantitative predictions for the specific heat and the compressibility of doped graphene sheets. We find that, at low temperatures, the specific heat has the usual normal-Fermi-liquid linear-in-temperature behavior, with a slope that 
is solely controlled by the renormalized quasiparticle velocity.
\end{abstract}

\submitto{\JPA}
\maketitle

\section{Introduction}

Graphene is a newly realized two-dimensional (2D) electron system 
that has attracted a great deal of interest in the scientific community because of the new physics which it exhibits and because of 
its potential as a new material for electronic technology~\cite{reviews,PT}.  The agent responsible for many of 
the interesting electronic properties of graphene sheets is the non-Bravais honeycomb-lattice
arrangement of Carbon atoms, which leads to a gapless semiconductor
with valence and conduction $\pi$-bands.
States near the Fermi energy of a graphene sheet are described by a spin-independent massless Dirac Hamiltonian~\cite{slonczewski} 
\begin{equation}\label{eq:h_kin}
{\cal H}_{\rm D} = v_{\rm F} {\bm \sigma}\cdot {\bm p}~,
\end{equation}
where $v_{\rm F}$ is the Fermi velocity, which is density-independent and roughly three-hundred times smaller that the velocity of light in vacuum, 
and ${\bm \sigma}=(\sigma^x,\sigma^y)$ 
is a vector constructed with two Pauli matrices $\{\sigma^i,i=x,y\}$, 
which operate on pseudospin (sublattice) degrees of freedom.
Note that the eigenstates of ${\cal H}_{\rm D}$ have a definite {\it chirality} rather than a definite pseudospin, {\it i.e.} they 
have a definite projection of the honeycomb-sublattice pseudospin onto the momentum ${\bm p}$.

When non-relativistic Coulombic electron-electron interactions
are added to the kinetic Hamiltonian (\ref{eq:h_kin}), graphene represents a new type of many-electron problem,
distinct from both an ordinary 2D electron gas (EG) and from quantum electrodynamics. 
The Dirac-like wave equation and the chirality of its eigenstates lead indeed to both unusual electron-electron interaction effects~\cite{diracgaspapers,barlas_prl_2007,polini_ssc_2007,polini_prb_2008,dassarmadgastheory} and to unusual response 
to external potentials~\cite{tomadin_prb_2008,rossi_prl_2008}. 

Within this low energy description, the properties of doped graphene sheets depend on the dimensionless coupling constant
\begin{equation}
\alpha_{\rm gr}= g \frac{e^2}{\epsilon \hbar v_{\rm F}}~,
\end{equation}
and on an ultraviolet cut-off $\Lambda = k_{\rm c}/k_{\rm F}$. 
Here $g = g_{\rm s} g_{\rm v}=4$ accounts for spin and valley degeneracy, 
$k_{\rm F}=(4\pi n/g)^{1/2}$ is the Fermi wave number with $n$ the electron density,
and $k_{\rm c}$ should be assigned a value corresponding to the
wavevector range over which the continuum model (\ref{eq:h_kin}) describes graphene.
For definiteness we take $k_{\rm c}$ to be such that $\pi k^2_{\rm c}=(2\pi)^2/{\cal A}_0$,
where ${\cal A}_0=3\sqrt{3} a^2_0/2$ is the area of the unit cell in the honeycomb lattice,
with $a_0 \simeq 1.42$~\AA~the Carbon-Carbon distance. With this choice
\begin{equation}
\Lambda = \frac{\sqrt{g}}{\sqrt{n{\cal A}_0}}~.
\end{equation}
The continuum model is useful 
when $k_{\rm c} \gg k_{\rm F}$, {\it i.e.} when $\Lambda \gg 1$. 

Vafek~\cite{vafek_prl_2007} has recently shown that the specific heat of undoped graphene sheets presents an anomalous low-temperature behavior showing a logarithmic suppression with respect to its noninteracting counterpart, $C_V(T \to 0)/C_V(0)\propto T/\ln(T)$. On the other hand, in Refs.~\cite{polini_ssc_2007,polini_prb_2008} we have demonstrated (see also Ref.~\cite{dassarmadgastheory}) that doped graphene sheets are normal (pseudochiral) Fermi liquids, with Landau parameters that possess, however, a quite distinct behavior from those of conventional 2D EGs. In this work we calculate the Helmholtz free energy ${\cal F}(T)$ of doped graphene sheets within the Random Phase Approximation (RPA)~\cite{Pines_and_Nozieres,Giuliani_and_Vignale}. This allows us to access important thermodynamic quantities, 
such as the compressibility and the specific heat, which can be calculated by taking appropriate derivatives of the free energy. We show  that, at low temperatures, the specific heat of doped graphene, contrary to the one of the undoped system~\cite{vafek_prl_2007}, has the usual linear-in-temperature behavior, which 
is solely controlled by the renormalized velocity of quasiparticles as in a normal Fermi liquid.

\section{The Helmoltz free energy and the Lindhard response function at finite temperature}

The free energy ${\cal F}={\cal F}_0+{\cal F}_{\rm int}$ is usually decomposed into the sum of a noninteracting term ${\cal F}_0$ and an interaction contribution ${\cal F}_{\rm int}$.
To evaluate the interaction contribution to the Helmholtz free energy
we follow a familiar strategy~\cite{Giuliani_and_Vignale} by combining
a coupling constant integration expression for ${\cal F}_{\rm int}$ valid for uniform continuum models ($\hbar=1$ from now on),
\begin{equation}\label{eq:int_free_energy}
{\cal F}_{\rm int}(T)=
\frac{N}{2}\int_{0}^{1}d\lambda\int \frac{d^2{\bm q}}{(2 \pi)^2}v_q\left[
S^{(\lambda)}(q, T) - 1 \right]~,
\end{equation}
with a fluctuation-dissipation-theorem (FDT) expression~\cite{Giuliani_and_Vignale} for the static structure factor,
\begin{equation}\label{eq:structurefactor}
S^{(\lambda)}(q, T) = -\frac{1}{\pi n}\int_0^{+\infty}
d\omega~\coth{(\beta\omega/2)}\Im m \chi^{(\lambda)}_{\rho\rho}(q,\omega, T)~.
\end{equation}
Here $v_q=2\pi e^2/(\epsilon q)$ is the 2D Fourier transform of the Coulomb potential and $\beta=(k_{\rm B} T)^{-1}$. We anticipate that this version of the FDT (in which the frequency integration has to be performed over the real-frequency axis) requires care in handling the plasmon contribution to ${\cal F}_{\rm int}(T)$ (see discussion below).

The RPA approximation for
${\cal F}_{\rm int}$ then follows from the RPA approximation for $\chi^{(\lambda)}_{\rho\rho}(q,\omega)$:
\begin{equation}\label{eq:chi_RPA}
\chi^{(\lambda)}_{\rho\rho}(q,\omega, T) =
\frac{\chi^{(0)}(q,\omega, T)}{1-\lambda v_q\chi^{(0)}(q,\omega, T)}
\end{equation}
where $\chi^{(0)}(q,\omega, T)$ is the noninteracting density-density response-function,
\begin{eqnarray}\label{eq:chi_0}
\chi^{(0)}(q,\omega,T) &=& g\lim_{\eta\to 0^+}\sum_{s,s'=\pm}\int
\frac{d^2{\bm k}}{(2\pi)^2}
\frac{1 + ss'\cos(\theta_{{\bm k}, {\bm k}+{\bm q}})}{2}\nonumber\\
&\times& \frac{n_{\rm F}(\varepsilon_{{\bm k}, s}) - n_{\rm F}(\varepsilon_{{\bm k}+{\bm q}, s'})}{\omega +
\varepsilon_{{\bm k}, s} - \varepsilon_{{\bm k}+{\bm q}, s'} + i\eta
}~.
\end{eqnarray}
Here $\varepsilon_{{\bm k}, s}=sv_{\rm F}k$ are the Dirac band energies and
$n_{\rm F}(\varepsilon)=\{\exp[\beta(\varepsilon - \mu_0)] + 1\}^{-1}$ is the usual
Fermi-Dirac distribution function, $\mu_0=\mu_0(T)$ being the noninteracting chemical potential. As usual, this is determined by the normalization condition
\begin{equation}
\mu_0(T) = \int_{-\infty}^{+\infty}d\varepsilon~\nu(\varepsilon)n_{\rm F}(\varepsilon)~,
\end{equation}
where $\nu(\varepsilon) = g \varepsilon/(2\pi v^2_{\rm F})$ is the noninteracting density of states. 
For $T \to 0$ one finds $\mu_0(T)=\varepsilon_{\rm F}-\pi^2 (T/T_{\rm F})^2/6$, 
where $T_{\rm F}=\varepsilon_{\rm F}/k_B$ is the Fermi temperature.
The factor in the first line of Eq.~(\ref{eq:chi_0}), which depends on the angle $\theta_{{\bm k},{\bm k}+{\bm q}}$ between ${\bm k}$ and ${\bm k}+{\bm q}$, describes the dependence of Coulomb scattering on the relative chirality $s s'$ of the interacting electrons.

\begin{figure}
\begin{center}
\includegraphics[width=0.60\linewidth]{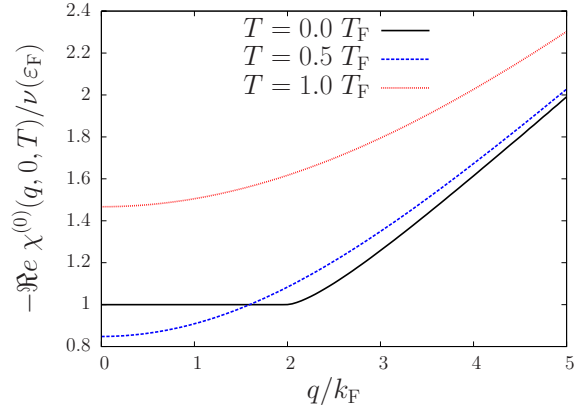}
\caption{The static response function $\Re e~\chi^{(0)}(q,0,T)$ [in units of $-\nu(\varepsilon_{\rm F})$] 
as a function of $q/k_{\rm F}$ for three values of $0 \leq T/T_{\rm F} \leq 1$.\label{fig:one}}
\end{center}
\end{figure}

\begin{figure}
\begin{center}
\includegraphics[width=0.45\linewidth]{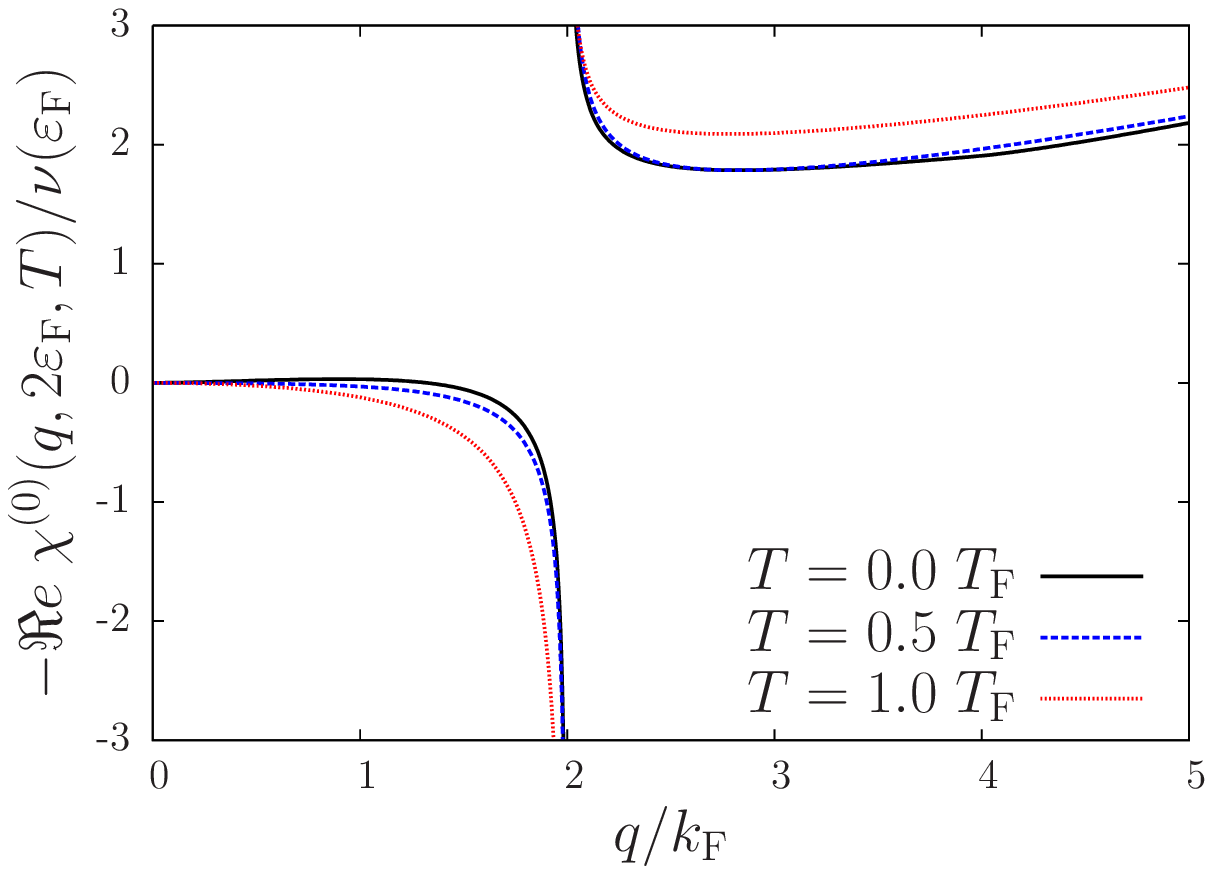}
\includegraphics[width=0.45\linewidth]{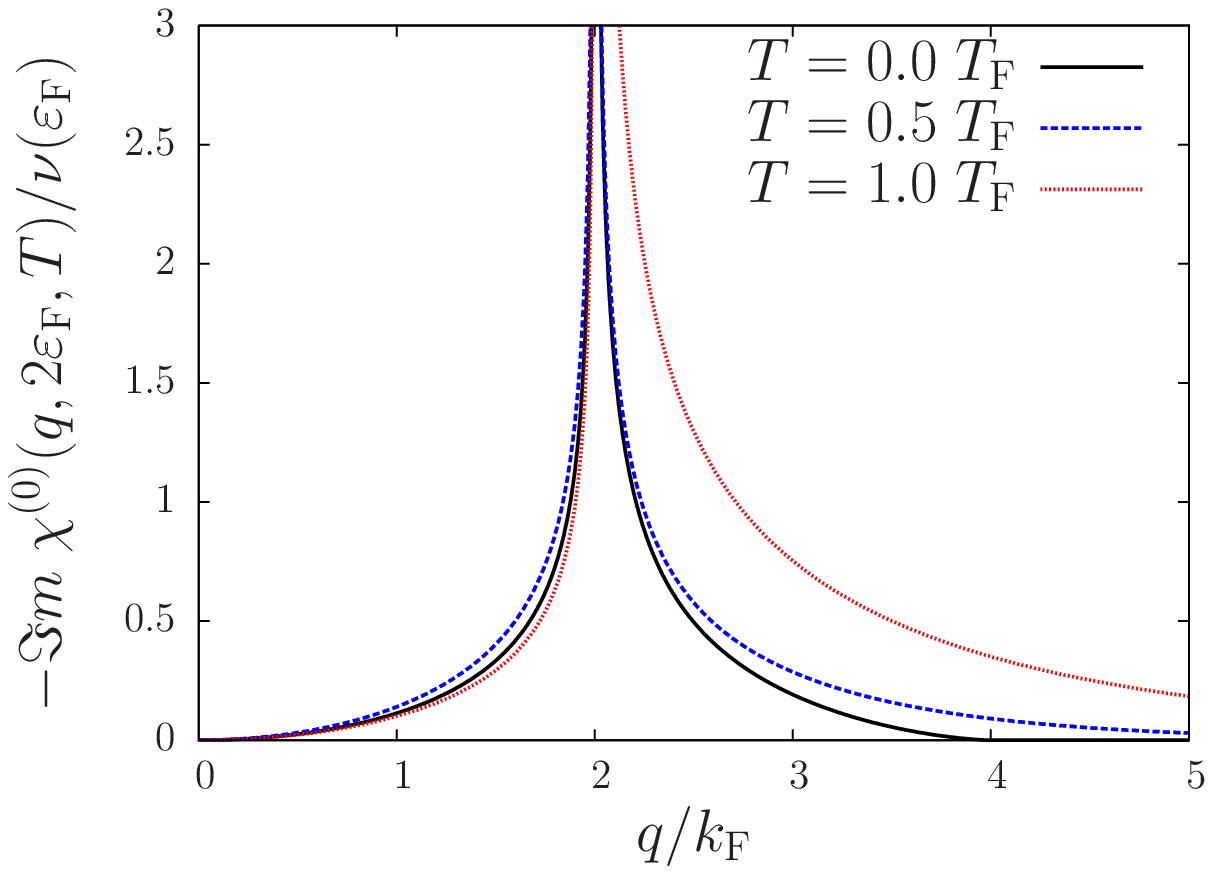}
\caption{Left panel: the real part of the dynamical response function $\Re e~\chi^{(0)}(q,\omega,T)$ 
[in units of $-\nu(\varepsilon_{\rm F})$] as a function of 
$q/k_{\rm F}$ for $\omega=2\varepsilon_{\rm F}$ and three values of $0 \leq T/T_{\rm F} \leq 1$. 
Right panel: same as in the left panel but for the imaginary part.\label{fig:two}}
\end{center}
\end{figure}

\begin{figure}
\begin{center}
\includegraphics[width=0.70\linewidth]{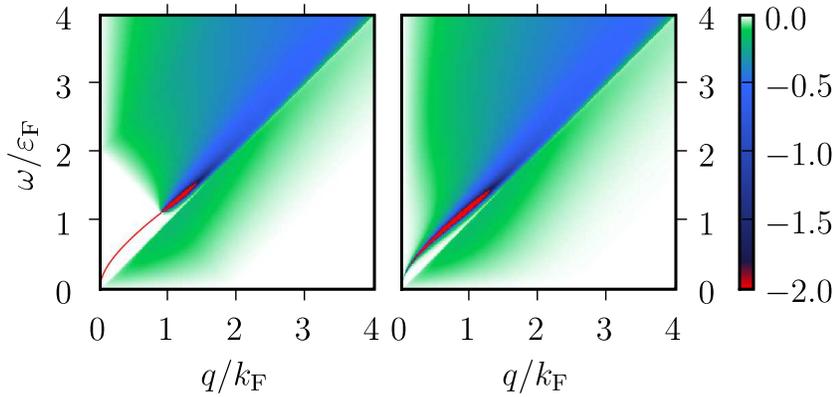}
\caption{Left panel: $\Im m~[\varepsilon^{-1}(q, \omega, T)]$ as a function of $q/k_{\rm F}$ and  $\omega/\varepsilon_{\rm F}$ for $\alpha_{\rm gr}=2$ and $T=0$. The red solid line is the plasmon dispersion relation. Right panel: same as in the left panel but for $T= 0.2~T_{\rm F}$ (corresponding roughly 
to room temperature).\label{fig:three}}
\end{center}
\end{figure}

After some straightforward algebraic manipulations we arrive at the following expressions for the imaginary [$\Im m~\chi^{(0)}(q,\omega,T)$] and the real [$\Re e~\chi^{(0)}(q,\omega,T)$] parts of the noninteracting 
density-density response function for $\omega>0$:
\begin{eqnarray}\label{eq:Im_chi_0}
\Im m~\chi^{(0)}(q,\omega,T)&=&
\frac{g}{4\pi}\sum_{\alpha = \pm}\Bigg\{\Theta(v_{\rm F} q-\omega)q^2 f(v_{\rm F}q,\omega)\nonumber\\
&\times&\left[G^{(\alpha)}_+(q,\omega,T)-G^{(\alpha)}_-(q,\omega,T)\right]\nonumber\\
&+&\Theta(\omega-v_{\rm F}q)q^2f(\omega,v_{\rm F}q)
\left[-\frac{\pi}{2}\delta_{\alpha,-}+H^{(\alpha)}_+(q,\omega,T)\right]\Bigg\}\nonumber\\
\end{eqnarray}
and
\begin{eqnarray}\label{eq:Re_chi_0}
\Re e~\chi^{(0)}(q,\omega,T)&=&
\frac{g}{4\pi}\sum_{\alpha = \pm}
\Bigg\{\frac{-2k_{\rm B}T\ln[1+e^{\alpha\mu_0/(k_{\rm B}T)}]}{v^2_{\rm F}}+
\Theta(\omega-v_{\rm F} q) \nonumber\\
&\times& q^2f(\omega,v_{\rm F}q)\left[G^{(\alpha)}_-(q,\omega,T)-G^{(\alpha)}_+(q,\omega,T)\right]\nonumber\\
&+&\Theta(v_{\rm F} q-\omega)q^2f(v_{\rm F}q,\omega)
\left[-\frac{\pi}{2}\delta_{\alpha,-}+H^{(\alpha)}_-(q,\omega,T)\right]\Bigg\}~.\nonumber\\
\end{eqnarray}
Here
\begin{equation}
f(x,y)= \frac{1}{2\sqrt{x^2-y^2}}~,
\end{equation}
\begin{equation}
G^{(\alpha)}_\pm(q,\omega,T)=\int_{1}^{\infty}du~\frac{\sqrt{u^2-1}}{\exp\left({\displaystyle \frac{|v_{\rm F} q u\pm\omega|-2\alpha\mu_0}{2k_{\rm B}T}}\right)+1}~,
\end{equation}
and
\begin{equation}
H^{(\alpha)}_\pm(q,\omega,T)=\int_{-1}^{1}du~
\frac{\sqrt{1-u^2}}{\exp\left({\displaystyle \frac{|v_{\rm F} q u\pm\omega|-2\alpha\mu_0}{2k_{\rm B}T}}\right)+1}~.
\end{equation}
These semi-analytical expressions for $\Re e~\chi^{(0)}(q,\omega,T)$ and $\Im m~\chi^{(0)}(q,\omega,T)$ constitute the first important result of this work. In Fig.~\ref{fig:one} we have plotted the static response, 
$\Re e~\chi^{(0)}(q,0,T)$, as a function of $q/k_{\rm F}$ for different values of $T/T_{\rm F}$. The temperature dependence of the Lindhard function at finite frequency is instead presented in Fig.~\ref{fig:two}. An illustrative 
plot of the imaginary part of the inverse RPA dielectric function
$\varepsilon(q, \omega, T)=1-v_q \chi^{(0)}(q,\omega,T)$ is reported in Fig.~\ref{fig:three}.

The coupling constant integration in Eq.~(\ref{eq:int_free_energy}) can be carried out partly analytically
due to the simple RPA expression (\ref{eq:chi_RPA}). We find that the interaction contribution to the free energy per particle $f_{\rm int}(T)$ is given by
\begin{eqnarray}\label{eq:int_free_energy_integrated}
f_{\rm int}(T) &\equiv& \frac{{\cal F}_{\rm int}(T)}{N}=\frac{1}{2}
\int\frac{d^2{\bm q}}{(2 \pi)^2}\left\{-\frac{1}{\pi n}\int_0^{+\infty}d\omega\coth{(\beta\omega/2)}\right.\nonumber\\
&\times&\left.\arctan\left[\frac{v_q\Im m ~\chi^{(0)}(q,\omega,T)}{1-v_q\Re e~\chi^{(0)}(q,\omega,T)}\right]-v_q\right\}\nonumber\\
&+&\frac{1}{2n}\int\frac{d^2{\bm q}}{(2 \pi)^2}\int_0^1 \frac{d\lambda}{\lambda}\coth{(\beta\omega_{\rm pl}/2)}
\Re e~\chi^{(0)}(q,\omega_{\rm pl},T)\nonumber\\
&\times&\left|{\frac{\partial[\Re e~\chi^{(0)}(q,\omega, T)]}{\partial \omega}}\right|^{-1}_{\omega = \omega_{\rm pl}}\,.
\end{eqnarray}
In this equation the first term comes from the smooth electron-hole contribution to 
$\Im m~\chi^{(\lambda)}_{\rho\rho}$, while the second 
term comes from the plasmon contribution; $\omega_{\rm pl}=\omega_{\rm pl}(q,T,\lambda)$ is the plasmon dispersion relation at coupling constant $\lambda$ which can be found numerically by solving the equation $1-\lambda v_q \Re e~\chi^{(0)}(q,\omega,T)=0$. Note that in a usual 2D EG 
the exchange energy starts to matter little for $T$ of order $T_{\rm F}$ because all the
occupation numbers are small and the Pauli exclusion principle matters little.  In the graphene
case however exchange interactions with the negative energy sea remain important as long as
$T$ is small compared to $v_{\rm F} k_{\rm c}/k_{\rm B} = T_{\rm F}\Lambda$. 

The free energy calculated according to Eq.~(\ref{eq:int_free_energy_integrated})
is divergent since it includes the interaction energy of the model's infinite sea of negative energy particles. 
Following Vafek~\cite{vafek_prl_2007},
we choose the free energy at $T=0$, $f(T=0)$,
as our ``reference" free energy, and thus introduce the regularized quantity
$\delta f\equiv f(T)-f(T=0)$. This again can be decomposed into the sum of a noninteracting contribution,
$\delta f_0(T \to 0)=-g\varepsilon_{\rm F}\pi^2 (T/T_{\rm F})^2/12$, and an interaction-induced contribution $\delta f_{\rm int}(T)=f_{\rm int}(T)-f_{\rm int}(T=0)$, which can be calculated from Eq.~(\ref{eq:int_free_energy_integrated}).  Numerical results for $\delta f_{\rm int}(T)$ as a function of the reduced temperature $T/T_{\rm F}$ are presented in the left panel of Fig.~\ref{fig:four}.

The low-temperature behavior of the interaction contribution to the free energy can be extracted analytically with some patience. 
After some lenghty but straightfoward algebra we find, to leading order in $\Lambda$,
\begin{eqnarray}\label{eq:crucial}
\delta f_{\rm int}(T\to 0)&=& \varepsilon_{\rm F} \frac{\pi^2}{3}
\left(\frac{T}{T_{\rm F}}\right)^2\frac{\alpha_{\rm gr}[1-\alpha_{\rm gr}\xi(\alpha_{\rm gr})]}{4g}~\ln{\Lambda}+{\rm R.~T.}~,
\end{eqnarray}
where the function $\xi(x)$, defined as in Eq.~(14) of Ref.~\cite{polini_ssc_2007}, is given by $\xi(x)=128/(\pi^2 x^3)- 32/(\pi^2 x^2)+ 1/x-h(\pi x/8)$, 
with
\begin{equation}
h(x)=\left\{
\begin{array}{ll}
{\displaystyle \frac{1}{2x^3\sqrt{1-x^2}}\arctan{\left(\frac{\sqrt{1-x^2}}{x}\right)}} & {\displaystyle {\rm for}~x<1}\vspace{0.1 cm}\\
{\displaystyle \frac{1}{4x^3\sqrt{x^2-1}}\ln{\left(\frac{x+\sqrt{x^2-1}}{x-\sqrt{x^2-1}}\right)}} & {\displaystyle {\rm for}~x>1}
\end{array}
\right.\,.
\end{equation}
The symbol ``${\rm R.~T.}$" in the l.f.s. of Eq.~(\ref{eq:crucial}) indicates ``regular terms", {\it i.e.} terms that, by definition, are finite in the limit $\Lambda \to \infty$.  Eq.~(\ref{eq:crucial}) represents the second important result of this work.

Befor concluding this Section, we remind the reader that in Ref.~\cite{polini_ssc_2007} it has been proven that the renormalized RPA 
quasiparticle velocity $v^\star$ is given, at weak coupling and to leading order in $\Lambda$, by
\begin{equation}\label{eq:velocity}
\frac{v^\star}{v_{\rm F}}=1+\frac{\alpha_{\rm gr}[1-\alpha_{\rm gr}\xi(\alpha_{\rm gr})]}{4g}~\ln{\Lambda}~.
\end{equation}

\section{The specific heat and the compressibility}

\begin{figure}
\begin{center}
\includegraphics[width=0.45\linewidth]{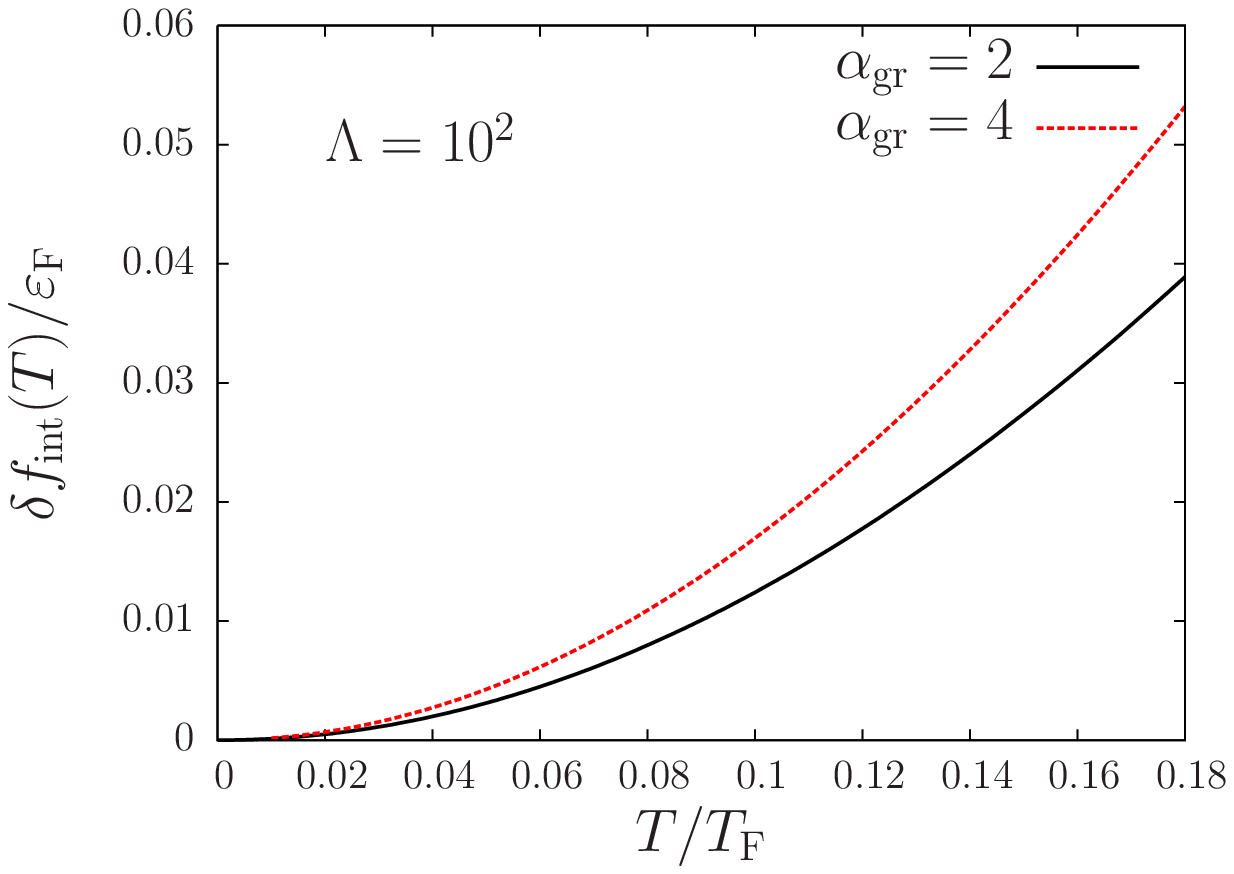}
\includegraphics[width=0.45\linewidth]{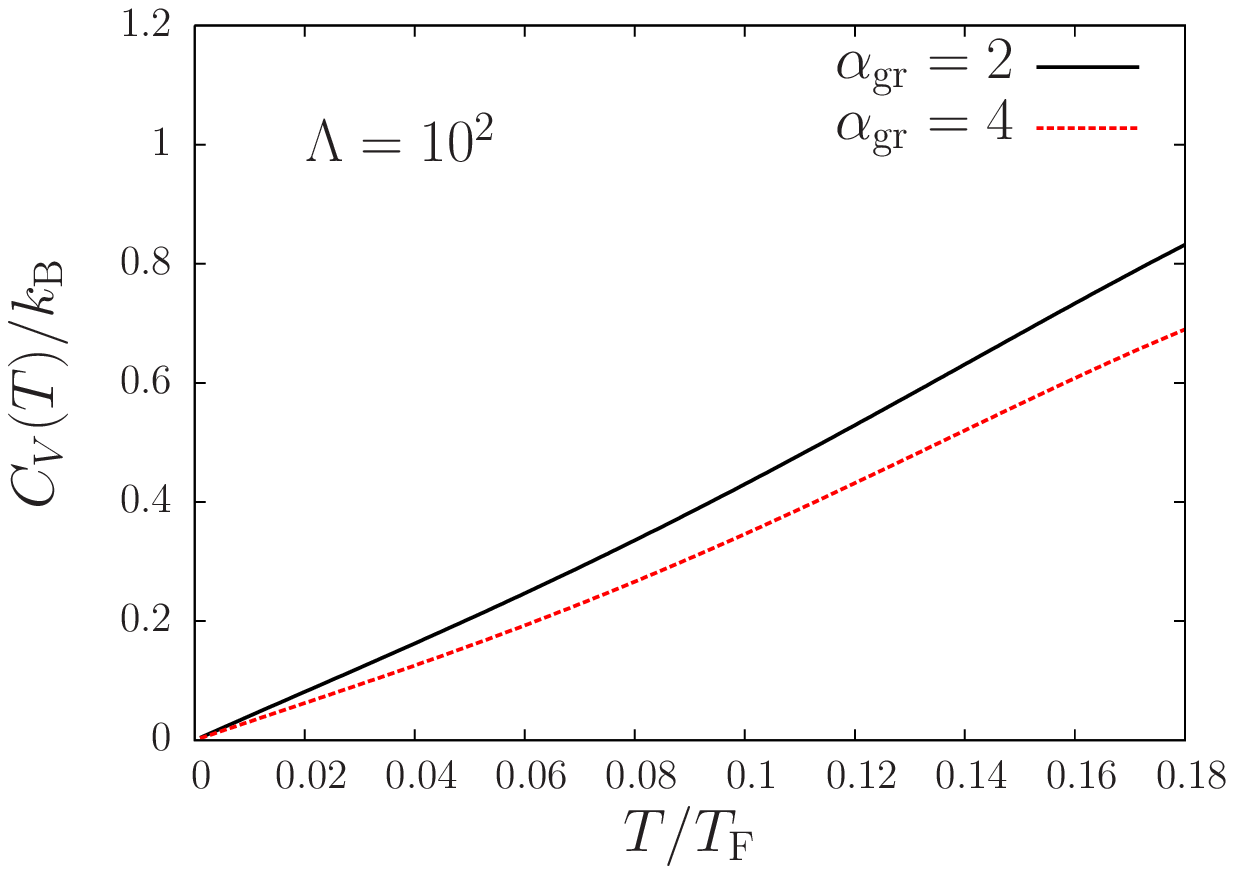}
\caption{Left panel: the (regularized) interaction contribution to the free energy $\delta f_{\rm int}(T)$ 
(in units of the Fermi energy $\varepsilon_{\rm F}$) as a function of $T/T_{\rm F}$ for $\Lambda=10^2$. 
Right panel: the specific heat $C_V(T)$ (in units of $k_{\rm B}$)
as a function of $T/T_{\rm F}$.\label{fig:four}}
\end{center}
\end{figure}

The specific heat can be calculated from the second derivative of the Helmholtz free energy,
$C_V=-T \partial^2 [n\delta f(T)]/\partial T^2$~\footnote{The second derivative is calculated using the full temperature dependent free-energy of the non-interacting system, $\delta f_0(T)$, and not its analytical expression reported above that is valid only for $T \ll T_{\rm F}$.}. Numerical results for $C_V(T)$ as a function of temperature are reported in Fig.~\ref{fig:four}.
We thus see that $\delta f_{\rm int}(T\to 0) \propto T^2$ in Eq.~(\ref{eq:crucial}) implies a conventional Fermi-liquid behavior with a linear-in-$T$ specific heat. Moreover, comparing Eq.~(\ref{eq:crucial}) with Eq.~(\ref{eq:velocity}) 
we find that the ratio between $C_V$ and its noninteracting value $C^{(0)}_V$ is given by
\begin{equation}
\lim_{T\to 0}\frac{C_V}{C^{(0)}_V}=\frac{v_{\rm F}}{v^\star}~,
\end{equation}
a well-known property of normal Fermi liquids~\cite{Pines_and_Nozieres,Giuliani_and_Vignale}. 
We are thus led to conclude, in full agreement with the zero-temperature calculations of the quasiparticle energy 
and lifetime performed in Refs.~\cite{polini_ssc_2007,polini_prb_2008}, that doped graphene sheets are normal Fermi liquids. Note that the fact that interactions enhance the quasiparticle velocity [see Eq.~(\ref{eq:velocity})] implies that the specific heat of doped graphene sheets is {\it suppressed} with respect to its noninteracting value.

\begin{figure}
\begin{center}
\includegraphics[width=0.6\linewidth]{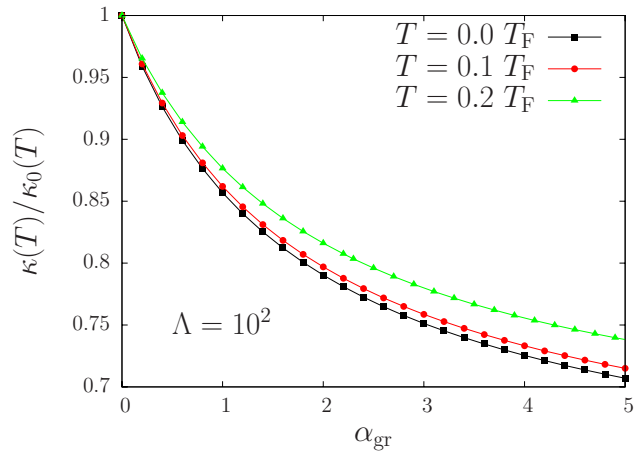}
\caption{The dimensionless ratio $\kappa(T)/\kappa_0(T)$ as a function of graphene's coupling constant $\alpha_{\rm gr}$ for three values of $0 \leq T/T_{\rm F} \leq 0.2$.\label{fig:five}}
\end{center}
\end{figure}

The compressibility can be calculated from the following equation
\begin{equation}\label{eq:compress}
\frac{1}{n^2\kappa(T)}=\frac{1}{n^2\kappa_0(T)}+\frac{\partial^2 [n \delta f_{\rm int}(T)]}{\partial n^2}~,
\end{equation}
where $\kappa^{-1}_0(T)$ is the inverse compressibility of the noninteracting system at finite temperature. 
In the low-temperature limit $1/[n^2\kappa_0(T\to 0)]=
n\varepsilon_{\rm F}/2+g n \varepsilon_{\rm F}\pi^2 (T/T_{\rm F})^2/48$. The dependence 
of the ratio $\kappa(T)/\kappa_0(T)$ on $\alpha_{\rm gr}$ and $T/T_{\rm F}$ is shown in Fig.~\ref{fig:five}.

\section{Conclusions}

In this work we have presented semi-analytical expressions for the real and the imaginary parts of the 
density-density linear-response function of noninteracting 
massless Dirac fermions at finite temperature. These results are very useful to study finite-temperature screening 
within the Random Phase Approximation. For example they can be used
to calculate the conductivity at finite temperature within Boltzmann transport theory and make quantitative comparisons with recent experimental results in unsuspended~\cite{morozov_prl_2008,chen_nature_nanotech_2008} and suspended graphene sheets~\cite{bolotin_prl_2008,du_nature_nanotech_2008}.

The Lindhard function at finite temperature is also extremely useful to calculate finite-temperature equilibrium properties of interacting massless Dirac fermions, such as the specific heat and the compressibility.
For example, in this work we have been able to show that, at low temperatures, the specific heat of interacting massless Dirac fermions has the usual normal-Fermi-liquid linear-in-temperature behavior, with a slope that 
is solely controlled by the renormalized quasiparticle velocity.

\ack

M.P. was partly supported by the CNR-INFM ``Seed Projects".

\end{document}